\documentclass[preprint,10pt]{elsarticle}

\usepackage{float}

\usepackage{siunitx}
\usepackage{booktabs}

\usepackage[hidelinks]{hyperref}

\usepackage[inline]{enumitem}
\setlist[enumerate]{label=(\arabic*), itemjoin={{, }}, itemjoin*={{, and }}, afterlabel=\unskip{{~}}}

\usepackage{multirow}
\usepackage{multicol}
\usepackage{adjustbox}
\usepackage{float}
\usepackage{makecell}
\usepackage{float}
\usepackage[ruled]{algorithm2e}
\usepackage{listings}

\lstdefinestyle{custom_java}{
  belowcaptionskip=1\baselineskip,
  basicstyle=\scriptsize,
  captionpos=b,
  breaklines=true,
  xleftmargin=\parindent,
  language=Java,
}

\usepackage{subcaption}
\usepackage{amsmath, amssymb, amsfonts}
\usepackage{graphicx}
\usepackage{textcomp}
\usepackage{xcolor}
\usepackage[capitalize,sort]{cleveref}

\usepackage{lineno}
\usepackage{csquotes}
\modulolinenumbers[5]

\newcommand{\includecode}[2][c]{\lstinputlisting[escapechar=, style=custom_#1]{#2}}

\usepackage{textcomp}

\usepackage{verbatim}

\journal{Journal of Systems and Software}
\biboptions{sort&compress}

\begin{document}
\nolinenumbers

\begin{frontmatter}

\title{A Pattern-based Approach to Detect and Improve Non-descriptive Test Names}

\author{Jianwei Wu}
\ead{wjwcis@udel.edu}

\author{James Clause}
\ead{clause@udel.edu}

\address{Dept. of Computer and Information Sciences\\University of Delaware\\Newark, DE, USA}

\begin{abstract}
Unit tests are an important artifact that supports the software development process in several ways. For example, when a test fails, its name can provide the first step towards understanding the purpose of the test. Unfortunately, unit tests often lack descriptive names. In this paper, we propose a new, pattern-based approach that can help developers improve the quality of test names of JUnit tests by making them more descriptive. It does this by detecting non-descriptive test names and in some cases, providing additional information about how the name can be improved. Our approach was assessed using an empirical evaluation on 34352 JUnit tests. The results of the evaluation show that the approach is feasible, accurate, and useful at discriminating descriptive and non-descriptive names with a 95\% true-positive rate.
\end{abstract}


\end{frontmatter}

\nolinenumbers

\section{Introduction}
\label{sec:intro}

Unit tests are an important artifact that supports the software development process in several ways.
In addition to helping developers ensure the quality of their software by checking for failures~\cite{daka2014survey}, they can also serve as an important source of documentation not only for human developers but also for automated software engineering tools (e.g., recent work on fault localization by \citeauthor{li2019deepfl} uses test name information~\cite{li2019deepfl}).
For example, when a test fails, its name can provide the first step towards understanding the purpose of the test and ultimately fixing the cause of the observed failure.
Similarly, a test's name can help developers decide whether a test should be left alone, modified, or removed in response to changes in the application under test and whether the test should be included in a regression test suite.

In this work, we believe that test names are \enquote{good} if they are descriptive (i.e., they accurately summarize both the scenario and the expected outcome of the test~\cite{trenk14}) and \enquote{bad} if they are not descriptive.
This is because descriptive names:
\begin{enumerate*}
\item make it easier to tell if some functionality is not being tested---if a behavior is not mentioned in the name of a test, then the behavior is not being tested
\item help prevent tests that are too large or contain unrelated assertions---if a test cannot be summarized, it likely should be split into multiple tests
\item serve as documentation for the class under test---a class's supported functionality can be identified by reading the names of its tests
\end{enumerate*}~\cite{zhang2015automatically}.

Unfortunately, unit tests often lack descriptive names.
For example, an exploratory study by \citeauthor{zhang2015automatically} found that only \SI{9}{\percent} of the \num{213423} test names they considered were complete (i.e., fully described the body of test) while \SI{62}{\percent} were missing some information and \SI{29}{\percent} contained no useful information (e.g., tests named \enquote{test})~\cite{zhang2015automatically}.
Poor test names can be due to developers writing non-descriptive or incomplete names.
They can also occur due to incomplete code modifications.
For example, a developer may modify a test's body but fail to make the corresponding changes to the test's name.
Regardless of the cause, non-descriptive test names complicate comprehension tasks and increase the costs and difficulty of software development.

Because non-descriptive names negatively impact software development, there have been several attempts to address this issue.
One approach has been to automatically generate names based on implementations (e.g.,~\cite{arcuri2014automated, zhang2015automatically, daka2017generating}).
For example, \citeauthor{zhang2015automatically} and \citeauthor{daka2017generating} use static and dynamic analysis, respectively, to extract important expressions from a test's body and natural language processing techniques to transform such expression into test names \cite{zhang2015automatically, daka2017generating}. 
While automatically generating names from bodies eliminates the possibility of mismatches between names and bodies the generated names do not always meet with developer approval (e.g., they may not fit with existing naming conventions).
Another approach is to help developers improve their existing names by suggesting improvements.  
For example, \citeauthor{host2009debugging} proposed an approach for Java methods and variables which uses a set of naming rules and related semantics~\cite{host2009debugging}, \citeauthor{li2019deepfl} provided a learning-based approach to locate software faults using test name information \cite{li2019deepfl}, and \citeauthor{allamanis2015suggesting} and \citeauthor{pradel2018deepbugs} use a model-based and a learning-based approach, respectively, to directly suggest better names or find name-based bugs to facilitate improvements~\cite{allamanis2015suggesting, pradel2018deepbugs}.

In this paper, we propose a new, pattern-based approach that can:
\begin{enumerate*}
\item detect non-descriptive test names by finding mismatches between the name and body of a given JUnit test
\item provide descriptive information that consists of the main motive of test, the property to be tested in the test, and the prerequisite needed in the test or the object to be tested (see~\cref{sec:test_patterns} for details) to facilitate the improvement of non-descriptive test names
\end{enumerate*}.
Unlike existing approaches that suggesting improvements, which were designed to handle general methods, our approach is specific to JUnit tests.
The narrower scope of the work allows it to take advantage of the highly repetitive structures that exist in both test names and bodies of JUnit tests (see~\cref{sec:test_patterns}).
From a high-level point of view, the approach uses a set of predefined patterns to extract descriptive information from both a test's name and body.
This information is then compared to find non-descriptive names (i.e., cases where the name does not accurately summarize the body).
When a mismatch is found, the information used by the approach can help developers address the mismatch and improve the quality of the test name.

To assess the pattern-based approach, we implemented it as an IntelliJ IDE plugin.
The plugin was then used to carry out an empirical evaluation of the quality of more than \num{34000} tests from \num{10} Java projects.
Overall, the results of our evaluation are promising and show that the pattern-based approach is feasible, accurate, and effective.

In particular, this work makes the following contributions:
\begin{itemize}
  \item A novel, pattern-based approach can detect non-descriptive test names of JUnit tests and provide descriptive information about the unit tests to help developers improve existing unit tests.
  \item A prototype implementation of the approach as an IntelliJ IDE plugin.
  \item An empirical evaluation on \num{10} Java projects that shows:
      \begin{enumerate*}
          \item the patterns are general and cover a majority of test names and bodies
          \item the patterns can accurately extract descriptive information from both test names and bodies
          \item the approach can accurately classify test names as either descriptive or non-descriptive
      \end{enumerate*}.
\end{itemize}

\section{Test Patterns}
\label{sec:test_patterns}

\begin{figure*}[t]
\centering
\begin{subfigure}{0.9\textwidth}
    \includecode[java]{tc_Sample.java}
    \caption{Try-catch statement.}
    \label{PatternExample_tc}
\end{subfigure}
\begin{subfigure}{0.9\textwidth}
    \includecode[java]{allassertion_sample.java}
    \caption{Single assertion.}
    \label{PatternExample_allA}
\end{subfigure}
\caption{Example test patterns.}
\label{fig:example-patterns}
\end{figure*}

We choose a pattern-based approach because unit tests often have similar structures that can be used to identify the purpose of a test from both its name and body.
More specifically, patterns can be used to extract:
\begin{enumerate*}
    \item the \textbf{action} which is the focus of the test (i.e., what the test is testing)
    \item the \textbf{predicate} which are the properties that will be checked by the test
    \item the \textbf{scenario} which are the conditions under which the action is being performed or the predicate is evaluated
\end{enumerate*}.

As examples of the common structures shared by unit test bodies, consider the code examples shown in \cref{fig:example-patterns}.
\Cref{PatternExample_tc} shows a unit test whose body consists of a \texttt{try-catch} statement.
The goal of this type of test is to perform the action under an optional scenario and then to check whether the action was successful or not.
In the test corpus used for pattern generation in \cref{tab:test-corpus}, we found more than \num{2800} tests (\SI{\approx 14}{\percent}) shared this structure.
Because of the regular structure of this type of test, it is possible to automatically extract its purpose in the form of its action, scenario, and predicate.
More specifically, the action is the \texttt{method invocation} that occurs before the \enquote{fail} statement in the \texttt{try} part of the \texttt{try-catch} statement and the scenario is the object on which the action is performed.
In \cref{PatternExample_tc}, the action of the test is \enquote{\textbf{execute}} and the scenario of the test is \enquote{\textbf{action}}, the object being tested.
For another example, \Cref{PatternExample_allA} presents a unit test whose body contains only a single assertion.
The goal of this type of test is to compare the result of an action under a required scenario to an expected predicate.
Again, this type of test is common: about \num{1000} tests in the corpus mentioned above (\SI{\approx 5}{\percent}) share this structure.
For this type of test, the action and the predicate can be found by looking at the actual (second) and expected (first) arguments to the assertion statement, respectively, and the scenario will again be the object on which the action is invoked.
Therefore, in \cref{PatternExample_allA}, the action of the test is \enquote{\textbf{entries}}, the predicate is \enquote{\textbf{getSampleElements}}, and the scenario is \enquote{\textbf{multimap}}.

Common patterns among test names can also be seen in the examples in \cref{fig:example-patterns}.
\Cref{PatternExample_tc} shows an example where the test name (ignoring the leading \enquote{test}) consists of a leading verb separated from the following noun by an underscore.
In our corpus, roughly \num{7000} (\SI{\approx 35}{\percent}) test names shared this structure.
For this structure, the action of the test name is the leading verb and the scenario is the following noun (i.e., in this example, the action is \enquote{Execute} and the scenario is \enquote{Action}).

Similarly, \cref{PatternExample_allA} that presents a test name that consists of a single word.
In our corpus, about \num{3400} (\SI{\approx 17}{\percent}) unit tests had one-word test names.
In this case, the action of the test name is simply the single word contained in the name (i.e., \enquote{Entries} is the action for this example).

In the remainder of this section, we explain the process we used to identify common test name and test body patterns and present a list of the patterns that we used to detect non-descriptive names (see~\cref{sec:approach}).

\subsection{Test Corpus}

\begin{table}[t]
\scriptsize
\centering
\caption{Considered projects for identifying test patterns.}
\begin{tabular}{
    l
    l
    S[table-format=5]
}
\toprule
\textbf{Project}             & \textbf{Commit Hash} & \textbf{\# Tests} \\
\midrule
Google Guava~\cite{guava}    & 473f8d2              & 14020             \\
JFreeChart~\cite{JFreeChart} & d03e68a              & 2176              \\
JaCoCo~\cite{jacoco}         & f0102f0              & 1323              \\
Weka~\cite{weka}             & d72b95e              & 436               \\
Barbecue~\cite{barbecue}     & 44a8632              & 154               \\
\midrule
\multicolumn{2}{r}{Total}                           & 18109             \\
\bottomrule
\end{tabular}
\label{tab:test-corpus}
\end{table}

To identify common patterns among unit test names and bodies we considered a set of \num{18109} tests comprised of the test suites from the \num{5} Java projects shown in \cref{tab:test-corpus}.
These projects are influential open-source projects taken from either Github~\cite{github} or SourceForge~\cite{sourceforge}.
Each project either has thousands of stars on Github (e.g., Google Guava) or has been downloaded more than \num{10000} times per week on SourceForge (e.g., Weka).
Each project focuses on a different domain: \enquote{Guava} is a general-purpose collection of utility classes, \enquote{JFreeChart} is a 2D chart library designed for Java applications, \enquote{JaCoCo} is a Java code coverage library often used in testing, \enquote{Weka} is a machine learning toolkit, and \enquote{Barbecue} is used for creating barcodes.
Moreover, they are written by different authors so their test suites are likely to have tests written in different ways.
Due to these criteria, the patterns we identify from these projects are likely to be general, rather than specific to any one test suite from a project or author.

\subsection{Test Body Patterns}
\label{sec:body-patterns}

\begin{table}[t]
\scriptsize
\centering
\caption{Test Body Patterns}
\begin{tabular}{l} 
\toprule
\textbf{Name} \\
\midrule
If Else  \\
Loop  \\
Try Catch  \\
Try Catch (Restricted) \\
Try Catch (Generalized) \\
All Assertion (Single) \\
All Assertion (Multiple) \\
Normal (Restricted) \\
Normal (Generalized) \\
No Assertion \\
No Assertion (Generalized) \\
No Assertion (Specialized for sole method) \\
No Assertion (Single declaration) \\
No Assertion (Single method invocation) \\
No Assertion (Single new object) \\
No Assertion (Multiple method invocations) \\
No Assertion (Multiple declarations) \\
\bottomrule
\end{tabular}
\label{tab:body-patterns}
\end{table}

\begin{figure}[t]
\centering
    \begin{subfigure}{0.9\textwidth}
        \begin{footnotesize}
\begin{verbatim}
@ITEM=3=}end
@ITEM=0=start{
@ITEM=7=try{
@ITEM=8=catch{
@ITEM=11=}try
@ITEM=10=}catch
@ITEM=15=fail
@ITEM=2=methodCall
@ITEM=-1=*

0 -1 7 -1 8 -1 10 -1 11 -1 3 -1
0 -1 7 -1 2 -1 15 -1 8 -1 10 -1 11 -1 3 -1

start{ * try{ * catch{ * }catch * }try * }end *  1847
start{ * try{ * methodCall * fail * catch{ * }catch * }try * }end *  1470
\end{verbatim}
        \end{footnotesize}
    \end{subfigure}
\caption{Mined Examples by ClaSP.}
\label{ClaSP_examples}
\end{figure}

To identify common body patterns, we used a semi-automated process based on applying frequent pattern mining to the statements contained in test bodies.
We chose to operate at the statement-level for two major reasons:
\begin{enumerate*}
    \item statements are the basic syntactic component of tests and standard unit tests are composed of statements~\cite{JUnitCook}
    \item while the entire test serves a purpose, individual statements encapsulate sub-steps towards achieving the overall goal~\cite{lakhotia1993understanding} such as the action, scenario, and predicate
\end{enumerate*}.

The first step in the process was to eliminate inconsequential differences (e.g., literals, variable names, etc.) by abstracting each statement to a number that encodes its type.
For example, declaration statements are assigned the number \num{1}, \texttt{method invocation} statements are assigned the number \num{2}, etc.
While more nuanced abstraction approaches are possible (e.g., def-use-based or graph-based), we found that this approach worked satisfactorily in practice.
We also added special symbols to explicitly encode the start and end of each test.
These markers are used later to filter out mined patterns that do not span entire tests.

The second step in the process was to apply the ClaSP frequent pattern mining algorithm~\cite{gomariz2013clasp} to the abstracted statements.
ClaSP is a novel algorithm that utilized vertical database strategy and heuristic to mine frequent closed patterns.
We chose to use ClaSP because it can efficiently mine the complete set of frequent, closed patterns from its input~\cite{fournier2017survey}.
This means that ClaSP ensures that each mined pattern has the highest frequency among its super-patterns (i.e., closed, similar to a class and its super-classes), and that long patterns, which are relevant for our purposes, can be mined efficiently.
To avoid confusion, we will refer to the output of ClaSP as \enquote{proto-patterns} as they serve as the basis for constructing our test body patterns.
As the output of this step, we generated \num{873} proto-patterns.

The final step in the process was to manually examine the proto-patterns to generate test body patterns.
This step is necessary because the proto-patterns contain duplicates, spurious entries (i.e., patterns that do not occur in the original tests), and patterns that do not span an entire test.
Additionally, since the proto-patterns may contain different setups of where the action, predicate, and scenario should be extracted, we wanted patterns that are both general and that allow for accurately extracting the action, predicate, and scenario.

In~\cref{ClaSP_examples}, the pattern mining process is clearly illustrated by a real-world example that is extracted during the process of pattern mining.
The example is composed of three parts:
\begin{enumerate*}
\item abstracting each statement to a number
\item using ClaSP to mine frequent patterns from the abstracted statements
\item manually examining the proto-patterns to generate test body patterns
\end{enumerate*}.
First, we utilized an automated script to convert statements to numbers to prepare the corpus for pattern mining, and each type of statement to number pair is also stored for reference (e.g., \texttt{methodCall} to \num{2}).
Second, after the mining results of ClaSP is completely generated, we collected all generated sequences (e.g.,~\num{0} \num{-1} \num{7} \num{-1} \num{8} \num{-1} \num{10} \num{-1} \num{11} \num{-1} \num{3} \num{-1}) as the proto-patterns and use the statement-number pair from the first step to reconstruct those patterns (e.g.,~$start\{ \star try\{ \star catch\{ \star \}catch \star \}try \star \}end \star \}$).
Last, we performed a manual examination of the proto-patterns to generate test body patterns.
From the last two lines (i.e., each of them is a reconstructed pattern with its number of matches) in~\cref{ClaSP_examples}, although both of them are mined patterns and the first one even has more matches (i.e., \num{1847}), we only selected the second one as one kind of representation of the \textit{Try Catch (Restricted)} body pattern shown in~\cref{tc_one} since the first one is a spurious entry.

Because the number of proto-patterns is large, we used various grouping strategies to merge similar proto-patterns.
In particular, we found that grouping by control-flow statements was effective as such statements often define the high-level structures of test bodies.
Another useful approach was to group the proto-patterns by common prefixes in order to identify statement types that were often repeated.
The resulting groups of proto-patterns were further examined to eliminate ones that did not include both the special start and end of test markers and ones that did not allow for identifying the action, scenario, and predicate.
Finally, the remaining proto-patterns were manually translated in to the \num{17} test body patterns shown in \cref{tab:body-patterns}.
For these selected proto-patterns, we manually examined each of them and extracted the action, predicate, and scenario from each pattern by reviewing matched test bodies from those considered projects in \cref{tab:test-corpus}.
In the remainder of this section each of these patterns will be described in more detail.

\begin{description}

\item[If Else]

\begin{figure}[t]
\centering
    \begin{subfigure}{0.6\textwidth}
        \includecode[java]{ifElse.java}
    \end{subfigure}
\caption{If Else.}
\label{ifElse}
\end{figure}

The motive behind \texttt{If Else} body pattern is to capture a type of test body that uses an if-else condition to fulfill its task by testing a particular \texttt{method invocation} under a given object in the if part, and use the \texttt{assertions} in the else part for its evaluation.
As shown in \Cref{ifElse}, the extracted action from this pattern is \enquote{$\langle action \rangle$} as the first \texttt{method invocation} in the \texttt{if} part of the \texttt{if-else} statement.
Then the extracted scenario under test will be the only \enquote{object} that is declared before the \texttt{if-else} statement as \enquote{$\langle scenario \rangle$}, and the \texttt{method invocation} positioned as the \enquote{actual} of the first \texttt{assertion} in the \texttt{else} part will be the \enquote{$\langle predicate \rangle$}.

\item[Loop]

\begin{figure}[t]
\centering
    \begin{subfigure}{0.65\textwidth}
        \includecode[java]{loopPattern.java}
    \end{subfigure}
\caption{Loop.}
\label{loopPattern}
\end{figure}

The motive of \textit{Loop} body pattern is to include any test body that is trying to repetitively test a \texttt{method invocation} under a specific \enquote{object}, and use its contained \texttt{assertion} to evaluate the outcomes.
The action in \Cref{loopPattern} is the first \texttt{method invocation} inside the \texttt{loop} as \enquote{$\langle action \rangle$}, the predicate of the test is the \texttt{method invocation} - \enquote{$\langle predicate \rangle$} positioned as \enquote{actual} part of the \texttt{assertion}, and the scenario of the test is the \enquote{object} used for the \texttt{loop} condition as \enquote{$\langle scenario \rangle$}.
In addition, the while \texttt{loop} is used here as an example, other types of \texttt{loop} are also supported.

\item[Try Catch]

\begin{figure}[t]
\centering
    \begin{subfigure}{0.65\textwidth}
        \includecode[java]{tryCP.java}
    \end{subfigure}
\caption{Try Catch.}
\label{tryCP}
\end{figure}

The motive of creating \textit{Try Catch} body pattern is to capture many test bodies that are trying to perform a \texttt{method invocation} under a required object and then to check whether the \texttt{method invocation} was successful or not.
Accordingly, the action of the body in \Cref{tryCP} is the \texttt{method} that was invoked as \enquote{$\langle action \rangle$}.
And the object used to invoke the \texttt{method} or the leading object declared outside the try-catch statement - \enquote{$\langle scenario \rangle$} will be the scenario of the test.
The \texttt{assertion} in the body is \emph{optional}, so there might be no predicate of the test.
If there is an \texttt{assertion}, then the predicate of the test is the \texttt{method invocation} - \enquote{$\langle predicate \rangle$} positioned in the \enquote{actual} part of the \texttt{assertion}.

\item[Try Catch (Restricted)] 

\begin{figure*}[t]
\centering
    \begin{subfigure}{0.8\textwidth}
        \includecode[java]{tc_one.java}
    \end{subfigure}
\caption{Try Catch (Restricted).}
\label{tc_one}
\end{figure*}

The motive of \textit{Try Catch (Restricted)} body pattern is to include a type of test body that is trying to perform a \texttt{method invocation} (i.e., action) under an optional object (i.e., scenario) and then to check if the \texttt{method invocation} was successfully performed.
Accordingly, the action of the body in \Cref{tc_one} is the method that was invoked as \enquote{$\langle action \rangle$}, and the object used to invoke the \texttt{method} - \enquote{$\langle scenario \rangle$} will be the scenario of the test but it is \emph{optional} for this pattern.
The \texttt{assertion} is also \emph{optional}, so there might be no predicate of the test.
If there is an \texttt{assertion}, then the predicate of the test is the \texttt{method invocation} - \enquote{$\langle predicate \rangle$} positioned in the \enquote{actual} part of the \texttt{assertion}.
As we mentioned in \Cref{PatternExample_tc}, that unit test is a standard match to the \enquote{Try Catch (Restricted)}.

\item[Try Catch (Generalized)] 

\begin{figure*}[t]
\centering
    \begin{subfigure}{0.65\textwidth}
        \includecode[java]{tc_any.java}
    \end{subfigure}
\caption{Try Catch (Generalized).}
\label{tc_any}
\end{figure*}

This pattern - \textit{Try Catch (Generalized)} body pattern is a more general form of the previous two types of \texttt{try-catch} statement-based body patterns.
Similarly, the motive of creating this pattern is to capture any test that is trying to perform a \texttt{method invocation} under a required object and to check if the \texttt{method invocation} was successful.
The action and the scenario of the body are in the same places as mentioned in the previous two patterns - \enquote{$\langle action \rangle$} and \enquote{$\langle scenario \rangle$} in \Cref{tc_any}.
Other statements might appear before the \enquote{$\langle scenario \rangle.\langle action \rangle$}, but they are considered as \enquote{setup} for the action and scenario.
The \texttt{assertion} for this pattern is still \emph{optional}, but it could appear in the catch part or outside the \texttt{try-catch} statement.
If there is an \texttt{assertion}, then the predicate of the test is the \texttt{method invocation} - \enquote{$\langle predicate \rangle$} positioned in the \enquote{actual} part of the \texttt{assertion}.

\item[All Assertion (Single)] 

\begin{figure*}[t]
\centering
    \begin{subfigure}{0.7\textwidth}
        \includecode[java]{AllA_single.java}
    \end{subfigure}
\caption{All Assertion (Single).}
\label{AllA_single}
\end{figure*}

A test body matched by the \textit{All Assertion (Single)} body pattern compares the result of an action under a required scenario to an expected predicate.
In \textit{All Assertion (Single)}, the single \texttt{assertion} contained in the test body is trying to compare different results, so the action is the \texttt{method invocation} placed in the \enquote{actual} position of the \texttt{assertion}.
The predicate of the test is the \texttt{method invocation} placed in the \enquote{expected} position of the \texttt{assertion}, and the scenario will be the \enquote{object} that invokes the action. 
Therefore, in \cref{AllA_single}, the action of the body is \enquote{$\langle action \rangle$}, the predicate is \enquote{$\langle predicate \rangle$} that is required for its comparison and the scenario is \enquote{$\langle scenario \rangle$}, which is also required to invoke the \enquote{$\langle action \rangle$}.
Like in \Cref{PatternExample_allA}, the unit test is a standard match to the \textit{All Assertion (Single)}.

\item[All Assertion (Multiple)] 

\begin{figure*}[t]
\centering
    \begin{subfigure}{0.8\textwidth}
        \includecode[java]{AllA_mutiple.java}
    \end{subfigure}
\caption{All Assertion (Multiple).}
\label{AllA_mutiple}
\end{figure*}

The \textit{All Assertion(Multiple)} body pattern serves as a more general form of the \textit{All Assertion (Single)} pattern.
The motive and the locations of action, predicate, and scenario are the same as the \textit{All Assertion (Single)} pattern.
There are two differences:
\begin{enumerate*}
    \item the test contains more than one \texttt{assertion} as long as they are testing the same action, predicate, or scenario.
    \item there is a new type of \texttt{nested method invocation} in the \texttt{assertion}
\end{enumerate*}.
In \cref{AllA_mutiple}, the action, predicate, and scenario for the first kind of \texttt{assertion} are the same as the \texttt{assertion} in \textit{All Assertion (Single)}.
For the second kind of \texttt{assertion}, the action of the body will be the outer \texttt{method invocation} as \enquote{$\langle action \rangle$} that is invoked by the scenario of the body as \enquote{$\langle scenario \rangle$}.
The inner \texttt{method invocation} - \enquote{$\langle predicate \rangle$} is the predicate of the body, and it serves as a further step of performing the main action.

\item[Normal (Restricted)] 

\begin{figure*}[t]
\centering
    \begin{subfigure}{0.675\textwidth}
        \includecode[java]{NP_23.java}
    \end{subfigure}
\caption{Normal (Restricted).}
\label{NP_2/3}
\end{figure*}

The motive of \textit{Normal (Restricted)} body pattern is to capture a type of test body that tries to perform an action under a specific scenario as its \enquote{setup} and evaluate it with a required predicate.
The action often appears in the initialization part of the leading \texttt{declaration}, but it can also be in the \enquote{actual} part of the only \texttt{assertion}.
The scenario is optional (i.e., none of the first two statements is \texttt{method invocation}), but it could be the first \texttt{method} being invoked before the final \texttt{assertion} or the \texttt{object} initialized in the leading \texttt{declaration}.
The predicate is the \texttt{method} name (e.g., \enquote{assertEquals}, \enquote{assertNotNull}, etc.) extracted from the \texttt{assertion}.
In \cref{NP_2/3}, the action of the body is \enquote{$\langle action \rangle$}, the predicate is \enquote{$\langle predicate \rangle$}, and the scenario is \enquote{$\langle scenario \rangle$}.

\item[Normal (Generalized)] 

\begin{figure}[t]
\centering
    \begin{subfigure}{0.675\textwidth}
        \includecode[java]{NP_any.java}
    \end{subfigure}
\caption{Normal (Generalized).}
\label{NP_any}
\end{figure}

The motive of \textit{Normal (Generalized)} body pattern is also to capture a type of test body that tries to perform an action under a specific scenario as its \enquote{setup} and evaluate it with a required predicate.
This pattern is an extended version of \textit{Normal (Restricted)}, so it shares the similar extraction of the action and predicate.
One major difference between this pattern and the previous one is that this pattern allows more statements to be included in the body, and another difference is that this pattern only considers the \texttt{method invocation} as the scenario since multiple objects can be declared in the test body.
In \cref{NP_any}, the action of the body is \enquote{$\langle action \rangle$}, the predicate is \enquote{$\langle predicate \rangle$}, and the scenario is \enquote{$\langle scenario \rangle$}.

\item[No Assertion] 

\begin{figure}[t]
\centering
    \begin{subfigure}{0.7\textwidth}
        \includecode[java]{NoAstP.java}
    \end{subfigure}
\caption{No Assertion.}
\label{NoAstP}
\end{figure}

From a structural perspective, the \textit{No Assertion} body pattern differs from other patterns due to its lack of any \texttt{assertion}, which is inspired by one of the test stereotypes mentioned in a recent work~\cite{li2018aiding}.
Nonetheless, we greatly extended this type of test body as the following patterns.
The motive of \textit{No Assertion} pattern is intended to perform an action under a required scenario, but there is often no primary predicate due to the lack of \texttt{assertion}.
However, this pattern requires at least three lines of code for its information extraction, and it attempts to extract the predicate of the test.
The primary position of the action is in the initialization part of the leading \texttt{declaration}, and it can be as the first \texttt{method} being invoked after the \texttt{declaration}.
The scenario is the object that invokes the first \texttt{method invocation} (i.e., the action) and is declared in the leading \texttt{declaration}.
The predicate is the secondary \texttt{method invocation} being invoked after the first \texttt{method invocation}.
In \cref{NoAstP}, the action of the body is \enquote{$\langle action \rangle$}, the predicate is \enquote{$\langle predicate \rangle$}, and the scenario is \enquote{$\langle scenario \rangle$}.

\item[No Assertion (Generalized)] 

\begin{figure}[t]
\centering
    \begin{subfigure}{0.7\textwidth}
        \includecode[java]{NoAstP_any.java}
    \end{subfigure}
\caption{No Assertion (Generalized).}
\label{NoAstP_any}
\end{figure}

The motive of \textit{No Assertion (Generalized)} body pattern is also intended to perform an action under a required scenario, but there is often no primary predicate due to the lack of \texttt{assertion}.
Because the required lines of code decrease to two lines, this pattern is capable of capturing more test bodies.
The action changes to the first \texttt{method} being invoked after the leading \texttt{declaration}.
The scenario is the \texttt{object} in the leading \texttt{declaration}.
In \cref{NoAstP_any}, the action of is \enquote{$\langle action \rangle$} and the scenario is \enquote{$\langle scenario \rangle$}.

\begin{figure}[t]
\centering
    \begin{subfigure}{0.45\textwidth}
        \includecode[java]{NoAstP_MC.java}
    \end{subfigure}
\caption{No Assertion (Single method invocation).}
\label{NoAstP_MC}
\end{figure}

\item[No Assertion (Specialized for sole method)]

\begin{figure}[t]
\centering
    \begin{subfigure}{0.6\textwidth}
        \includecode[java]{NoAstP_one.java}
    \end{subfigure}
\caption{No Assertion (Specialized for sole method).}
\label{NoAstP_one}
\end{figure}

\begin{figure*}[t]
\centering
    \begin{subfigure}{0.7\textwidth}
        \includecode[java]{NoAstP_SD.java}
    \end{subfigure}
\caption{No Assertion (Single declaration).}
\label{NoAstP_SD}
\end{figure*}

The \textit{No Assertion (Specialized for sole method)} body pattern is a distinctive kind of no assertion-based pattern.
The specialization of this pattern is that it only captures a test body with only one \texttt{method invocation} across all its statements, which could be an independent \texttt{method invocation} or an argument from any statement in the body.
In \cref{NoAstP_one}, the motive of this pattern is to perform the sole action (i.e., the \texttt{method invocation}) under a required scenario, so the action of the body is \enquote{$\langle action \rangle$} and the scenario of the body is \enquote{$\langle scenario \rangle$}.

\item[No Assertion (Single declaration)]

Creating \textit{No Assertion (Single declaration)} body pattern is to include a special kind of no assertion body pattern that can capture any test body with a sole declaration.
The motive of this pattern is testing an \enquote{object} that is initialized by a required \texttt{method} to check if the \enquote{object} can be successfully initialized.
In \cref{NoAstP_SD}, the action of the body is \enquote{$\langle action \rangle$} as the \texttt{method} being invoked, and the scenario of the body is \enquote{$\langle scenario \rangle$} as the object being tested.

\item[No Assertion (Single method invocation)]

Creating \textit{No Assertion (Single method invocation)} body pattern is to include a special kind of no assertion body pattern that can capture any test body with a sole \texttt{method invocation}.
The motive of this pattern is performing an action that is under a required argument (i.e., predicate) to check if the action can be successfully performed.
In \cref{NoAstP_MC}, the action of the body is \enquote{$\langle action \rangle$} as the \texttt{method} being invoked, and the predicate of the body is the \enquote{$\langle predicate \rangle$} as the inner argument of the \texttt{method invocation}.

\item[No Assertion (Single new object)]

\begin{figure}[t]
\centering
    \begin{subfigure}{0.75\textwidth}
        \includecode[java]{NoAstP_new.java}
    \end{subfigure}
\caption{No Assertion (Single new object).}
\label{NoAstP_new}
\end{figure}

The \textit{No Assertion (Single new object)} body pattern is also a special kind of no assertion body pattern that can capture any test body with a sole new object initialization.
The motive of this pattern is initializing a \texttt{new object} that is chained to two required \texttt{methods} to check if the \texttt{new object} can be successfully initialized.
In \cref{NoAstP_new}, the action of the body is \enquote{$\langle action \rangle$} as the last \texttt{method} being invoked, the predicate is the \enquote{$\langle predicate \rangle$} as the first \texttt{method} being invoked, and the scenario is \enquote{$\langle scenario \rangle$} as the new object being initialized.

\begin{figure}[H]
\centering
    \begin{subfigure}{0.7\textwidth}
        \includecode[java]{NoAstP_m_dec.java}
    \end{subfigure}
\caption{No Assertion (Multiple declarations).}
\label{NoAstP_m_dec}
\end{figure}

\begin{figure}[H]
\centering
    \begin{subfigure}{0.45\textwidth}
        \includecode[java]{NoAstP_m_MC.java}
    \end{subfigure}
\caption{No Assertion (Multiple method invocations).}
\label{NoAstP_m_MC}
\end{figure}

\item[No Assertion (Multiple declarations)]

The \textit{No Assertion (Multiple declarations)} body pattern is to create an extension of the \textit{No Assertion (Single declaration)} pattern, and it allows more than one line of code in any captured test body.
In \cref{NoAstP_m_dec}, the motive and information extraction are the same as the \enquote{single method} version: the action of the body is \enquote{$\langle action \rangle$} as the \texttt{method} being invoked, and the scenario is \enquote{$\langle scenario \rangle$} as the object being tested.
Also, the scenario in this pattern needs to be the most frequently evaluated object in the test body, and the action is served as a required argument of that object.

\item[No Assertion (Multiple method invocations)]

The \textit{No Assertion (Multiple method invocations)} body pattern is to create an extension of the \textit{No Assertion (Single method invocation)} pattern, and it allows more than one line of code in any captured test body.
The motive and the information extraction are the same as the \enquote{single method} version: the action of the body is \enquote{$\langle action \rangle$} as the \texttt{method} being invoked, and the predicate is the \enquote{$\langle predicate \rangle$} as the inner argument of the \texttt{method invocation}, which is also shown in \cref{NoAstP_m_MC}.

\begin{figure}[H]
\centering
    \begin{subfigure}{0.65\textwidth}
        \includecode[java]{divided-dual-verb-n.java}
    \end{subfigure}
\caption{Divided Duel Verb Phrase.}
\label{fig:divided-dual-verb}
\end{figure}

\begin{figure}[H]
\centering
    \begin{subfigure}{0.65\textwidth}
        \includecode[java]{Is_Past-n.java}
    \end{subfigure}
\caption{Is And Past Participle Phrase.}
\label{fig:IsAndPast}
\end{figure}

\begin{figure}[H]
\centering
    \begin{subfigure}{0.65\textwidth}
        \includecode[java]{vmn-n.java}
    \end{subfigure}
\caption{Verb With Multiple Nouns Phrase.}
\label{fig:name_vmn}
\end{figure}

Also, the action in this pattern needs to be the most frequently invoked \texttt{method} in the test body, and the predicate is associated with the action as its inner argument.

\end{description}

\begin{table}[t]
\scriptsize
\centering
\caption{Test Name Patterns.}
 \begin{tabular}{l} 
 \toprule
 \textbf{Name} \\
 \midrule
 Verb With Multiple Nouns Phrase \\
 Divided Duel Verb Phrase \\
 Is And Past Participle Phrase \\
 Try Catch \\
 Duel Verb Phrase \\
 Noun Phrase \\
 Single Entity \\
 Verb Phrase Without Prepended Test \\
 Verb Phrase With Prepended Test \\
 Regex Match \\
 \bottomrule
\end{tabular}
\label{tab:name-patterns}
\end{table}

\subsection{Test Name Patterns}
\label{sec:name-patterns}

Because test names are easier to compare since they are shorter than test body we were able to use a fully manual process for identifying commonalities among test names.
We found that test names typically fall into two main categories: names that have a common structural format and names that have a common grammatical structure.
For the first category, regular expressions can be used directly on the test names to identify relevant pieces of information.
For the second category, additional information such as the part of speech of each word in the test name is needed.
To obtain this information we used an approach recommended by \citeauthor{olney16tagging}:
\begin{enumerate*}
\item convert each test name to a sentence by using a purpose-built identifier splitter and prepending the result with the word \enquote{I}
\item apply the Stanford Tagger~\cite{StanfordTagger}
\end{enumerate*}~\cite{olney16tagging}.
The resulting \num{10} name patterns are shown in \cref{tab:name-patterns} and each is described with more details in the remainder of the section.

\begin{description}

\item[Verb With Multiple Nouns Phrase]

This name pattern aims to match a test name that is composed of a prefix of \enquote{test} and a verb phrase with multiple nouns.
In \cref{fig:name_vmn}, the first word after the leading \enquote{test} should be tagged as \enquote{verb} that is the action of the name, and the following three \enquote{nouns} are combined as the scenario of the name.

\item[Divided Duel Verb Phrase]

\begin{figure}[t]
\centering
    \begin{subfigure}{0.65\textwidth}
        \includecode[java]{try-catch-n.java}
    \end{subfigure}
\caption{Try Catch (Name).}
\label{fig:try-catch}
\end{figure}

\begin{figure}[t]
\centering
    \begin{subfigure}{0.65\textwidth}
        \includecode[java]{dual-verb-n.java}
    \end{subfigure}
\caption{Duel Verb Phrase.}
\label{fig:dual-verb}
\end{figure}

The motive of this name pattern is to match a type of test names that has a leading "test" followed by a \enquote{verb-noun-verb-noun} structure.
In \cref{fig:divided-dual-verb}, the first word tagged as \enquote{verb} is the action of the name, and the second word tagged as \enquote{noun} is the scenario of the name that is same as the fourth word.
The third word should be tagged as \enquote{verb} that is the scenario of the name.

\item[Is And Past Participle Phrase]

This name pattern is intended to match any test name that has a \enquote{verb-verb} structure, and the second \enquote{verb} should be in its past participle form.
In \cref{fig:IsAndPast}, the first \enquote{verb} is the action of the name, and the second \enquote{verb} is the predicate of the name.
If there is a following \enquote{noun} after the second \enquote{verb}, it will be the scenario of the name.
However, there were not enough pattern matches to support the scenario, so it is currently not included in this name pattern.

\item[Try Catch (Name)]

The \enquote{Try Catch} name pattern is designed for test names, and this name pattern belongs to the regular expression-based name patterns.
\Cref{fig:try-catch} shows a more representative sub-pattern of this pattern than other sub-patterns.
In the figure, the action of the name is placed before the divider - \enquote{Throws}, and the predicate of the name is placed after the divider.
Moreover, this name pattern is often related to a \texttt{try-catch} condition that will be tested in the test.

\item[Duel Verb Phrase]

\begin{figure}[t]
\centering
    \begin{subfigure}{0.65\textwidth}
        \includecode[java]{verb-without-test-n.java}
    \end{subfigure}
\caption{Verb Phrase Without Prepended Test.}
\label{fig:verb-without-test}
\end{figure}

This name pattern aims to match a type of test names that has a \enquote{verb-verb-noun} structure.
In \cref{fig:dual-verb}, the action of the name is the first word tagged as \enquote{verb}, the predicate is the second word also tagged as \enquote{verb}, and the scenario is the third word tagged as \enquote{noun}.

\begin{figure}[H]
\centering
    \begin{subfigure}{0.65\textwidth}
        \includecode[java]{verb-with-test-n.java}
    \end{subfigure}
\caption{Verb Phrase With Prepended Test.}
\label{fig:verb-with-test}
\end{figure}

\begin{figure}[H]
\centering
     \begin{subfigure}{0.65\textwidth}
        \includecode[java]{noun-phase-n.java}
    \end{subfigure}
\caption{Noun Phrase.}
 \label{fig:noun-phrase}
\end{figure}

\begin{figure}[H]
\centering
     \begin{subfigure}{0.65\textwidth}
        \includecode[java]{single-entity-n.java}
    \end{subfigure}
\caption{Single Entity.}
\label{fig:single-entity}
\end{figure}

\item[Noun Phrase]

This name pattern is set to match any test name that only has one word tagged as a \enquote{noun}.
As shown in \cref{fig:noun-phrase}, the only \enquote{noun} is the scenario of the name, and there is often no action or predicate in the name.

\item[Single Entity]

The \enquote{Single Entity} name pattern also belongs to the regular expression-based name patterns, and a representative sub-pattern is shown in \cref{fig:single-entity}.
After the leading \enquote{test}, the combination of all following words is the action of the name.
Nonetheless, the action of the name needs to fulfill a special requirement that requires the action to be matched to one of the \enquote{method under test}~\cite{zhang2016towards}.
A \enquote{method under test} is a \texttt{method} that is being tested in the test or the test class.
When the action of the name is matched to a \enquote{method under test} (i.e., identical names), it will be counted as a pattern match to this name pattern.

\item[Verb Phrase Without Prepended Test]

This name pattern aims to match any test name that is a \enquote{verb phrase} without a prepended word - \enquote{test}.
The \enquote{verb phrase} consists of a leading \enquote{verb} with a following \enquote{noun}, and there is a secondary \enquote{verb} (i.e., optional) that comes after the \enquote{noun}.
In \cref{fig:verb-without-test}, the action of the name is the leading \enquote{verb}, the predicate of the name is the secondary \enquote{verb}, and the scenario of the name is the \enquote{noun} between the action and the predicate.

\item[Verb Phrase With Prepended Test]

This name pattern aims to match any test name that is a \enquote{verb phrase} with a prepended word - \enquote{test}.
The \enquote{verb phrase} also consists of a leading \enquote{verb} with a following \enquote{noun}, and there is a secondary \enquote{verb} (i.e., optional) that comes after the \enquote{noun}.
In \cref{fig:verb-with-test}, the action of the name is the leading \enquote{verb}, the predicate of the name is the secondary \enquote{verb}, and the scenario of the name is the \enquote{noun} between the action and the predicate.

\begin{figure}[t]
\centering
    \begin{subfigure}{0.675\textwidth}
        \includecode[java]{regex-match-n.java}
    \end{subfigure}
\caption{Regex Match.}
\label{fig:regex-match}
\end{figure}

\begin{figure*}[t]
  \centering
  \includegraphics[scale=0.25]{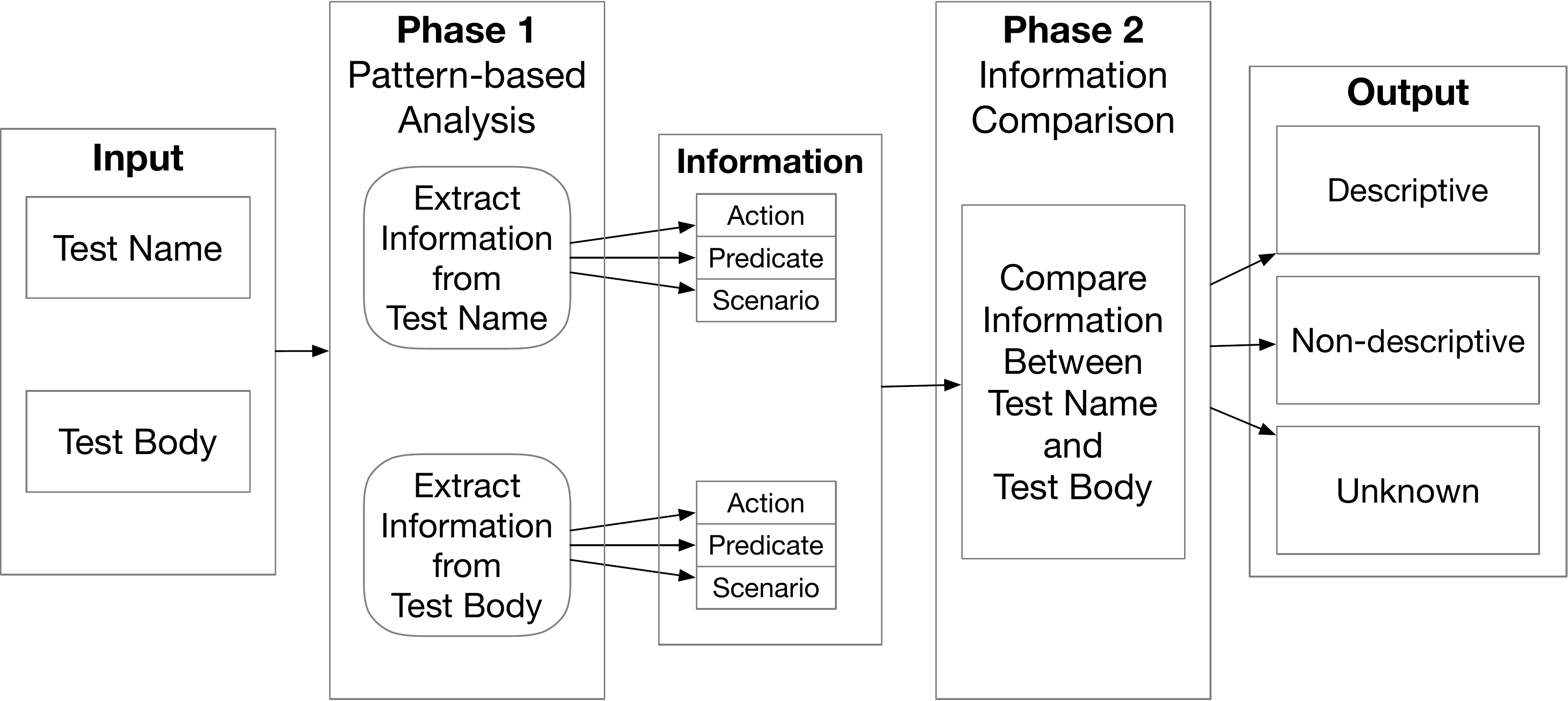}
  \caption{Overview of the pattern-based approach.}
  \label{fig:approach}
\end{figure*}

\item[Regex Match]

This name pattern is a collection of \num{70} regular expression-based sub-patterns.
For example, three of the most representative sub-patterns are shown in \cref{fig:regex-match}.
The first two sub-patterns show a special condition that need to perform the predicate under the defined scenario or execute the action after the scenario is performed.
The last sub-pattern is to execute the predicate while the right scenario is successfully performed, and a pattern match in practice is shown in \cref{PatternExample_detectedPoorName_code}.

\end{description}

\section{A Pattern-based Approach to Detect Non-descriptive Test Names}
\label{sec:approach}

\Cref{fig:approach} presents a high-level overview of our pattern-based approach for detecting non-descriptive test names.
As the figure shows, the approach takes as input a unit test comprised of its name and body.
It then assesses the descriptiveness of the test's name using two phases.
The first phase, \emph{pattern-based analysis}, uses the test patterns described in \cref{sec:test_patterns} to extract descriptive information from both the test name and the test body.
The second phase, \emph{information comparison}, compares the descriptive information extracted from the name and body against each other.
This information comparison process allows for not only detecting non-descriptive test names (i.e., mismatches between the information), but also in some cases indicating to developers how the name could be improved.
The remainder of this section describes the two steps of the approach in more detail.

\subsection{Phase 1: Pattern-based Analysis}
\label{sec:Pattern-based-Analysis}

\begin{figure}[t]
    \centering
    \begin{subfigure}{0.9\textwidth}
        \includecode[java]{approach-example.java}
        \caption{Example test that is matched by more than one name pattern and more than one body pattern.}
        \label{fig:approach-0}
    \end{subfigure}
    \\[1ex]
        \begin{subfigure}[b]{\columnwidth}
            \centering
            \begin{tabular}{rcc}
            \toprule
                      & Single Entity  & Verb Phrase With Prepended Test \\
            \midrule
            action    & GetSSLProtocol & Get \\
            predicate & ---            & --- \\
            scenario  & ---            & SSL \\
            \bottomrule
            \end{tabular}
            \caption{Comparison of information extracted by both matching name patterns for the test shown in \cref{fig:approach-0}.}
            \label{fig:approach-1}
        \end{subfigure}
    \\[1ex]
    \begin{subfigure}[b]{\columnwidth}
        \centering
        \begin{tabular}{rcc}
        \toprule
                  & Normal (Restricted) & Normal (Generalized) \\
        \midrule
        action    & getSSLProtocol    & getSSLProtocol \\
        predicate & assertNotNull     & assertNotNull \\
        scenario  & protocol          & --- \\
        \bottomrule
        \end{tabular}
        \caption{Comparison of information extracted by both matching body patterns for the test shown in \cref{fig:approach-0}.}
        \label{fig:approach-2}
    \end{subfigure}    
    \caption{Example to illustrate the ordering patterns is necessary.}
    \label{fig:approach-example}
\end{figure}

The first phase of the approach is relatively straightforward as it consists mainly of applying the patterns described in \cref{sec:name-patterns,sec:body-patterns} to the provided test name and body.
If a pattern matches against a name or body, the values it extracts as the action, predicate, and scenario are passed as input to the second stage.
If none of the name patterns match or none of the body patterns match, empty values are passed instead.

Generally, if a name or body meets all requirements of a name\slash body pattern, the name or body is counted as a match to that name\slash body pattern.
In~\cref{sec:test_patterns}, the requirements of matching each test pattern is stated in a pattern-by-pattern style.
For an example of how to match the name pattern, the \enquote{Noun Phrase} name pattern can be matched to a test name that is only composed of a leading \enquote{test} and an ending noun (i.e., requirements are fulfilled, and the name is considered to be a match to the \enquote{Noun Phrase} name pattern), and the ending noun is extracted from the name as the scenario of the name.
For an example of how to match the body pattern, the \enquote{All Assertion (Single)} body pattern can be matched to a test body that only contains a single and complete JUnit \texttt{assertion} (i.e., requirements are fulfilled, and the body is considered to be a match to the \enquote{All Assertion (Single)} body pattern).
The \texttt{expected} part of the \texttt{assertion} is extracted from the body as the predicate of the body, and the \texttt{actual} part of the \texttt{assertion} should be a complete \texttt{method invocation} that contains an \texttt{object} and a \texttt{method call}.
The \texttt{object} in the \texttt{method invocation} is extracted as the scenario of the body, and the \texttt{method call} is extracted as the action of the body.

After a match is found, the matched name\slash body pattern can extract the components from the name\slash body by using the corresponding positions of the action, predicate, and scenario.
Similar to the two examples of matching a pattern to a test name or body that we already mentioned, the extraction process is also straightforward.
For the same example of the \enquote{Noun Phrase} name pattern, the approach can automatically parse the test name with part-of-speech tags and stores every word in the name with its original order and part-of-speech tag.
Then the approach first rules out irrelevant test names (e.g., test names that contain more than two words) and then extracts the first and only noun in the name to be the scenario of the name
For the same example of the \enquote{All Assertion (Single)} body pattern, the approach is also able to automatically parse code from the statement-level and identifies different types of statements.
Therefore, the approach first rules out any test body with more than one statement or contains any kind of statement other than JUnit \texttt{assertion}, and it then parse the JUnit \texttt{assertion} to extract the \texttt{expected} part and the \texttt{actual} part.
After all parts of the \texttt{assertion} are gathered, the approach extracts the \texttt{expected} part (i.e., which is often a \texttt{method invocation}) as the predicate of the body, the \texttt{object} in the \texttt{actual} part as the scenario of the body, and the \texttt{method call} in the \texttt{actual} part as the action of the body.
When every component is successfully extracted from both the test name and body, the approach will determine if the name is descriptive or non-description and generate a report for the associated test as shown~\cref{sec:InformationComparison}.

The main complexity in this phase arises from the fact that more than one pattern may match a name or body.
For example, \cref{fig:approach-0} shows a unit test that can be matched by more than one pattern.
More specifically, the test's body can be matched by both the restricted and generalized versions of the \enquote{Normal} pattern and the test's name can be matched by both the \enquote{Single Entity} and \enquote{Verb Phrase With Prepended Test} patterns.

While more than one pattern may match the same name or body, there is often one pattern that is preferred either because it is more accurate at extracting information or it can extract more information.
For example, the difference in information extracted by matching patterns can be seen in \cref{fig:approach-1,fig:approach-2}.
Each of these figures, the rows show the values extracted as the action, predicate, and scenario for the patterns shown in the corresponding columns.
A dash (---) indicates an empty value that occurs when a pattern did not extract a value for the corresponding type of information.
\Cref{fig:approach-1} is an example of when one pattern may be more accurate at extracting information.
In this case, the \enquote{Single Entity} pattern correctly extracts \enquote{GetSSLProtocol} as the action and does not extract a value for the predicate or scenario while the \enquote{Verb Phrase With Prepended Test} pattern incorrectly identifies the action and scenario (i.e., \enquote{Get} is tagged as \enquote{verb} for the action and \enquote{SSL} is tagged as \enquote{noun} for the scenario).
\Cref{fig:approach-2} is an example of when one pattern may extract more information.
In this case, both the \enquote{Normal (Restricted)} and \enquote{Normal (Generalized)} patterns correctly identify the action as \enquote{getSslProtocol} and the predicate as \enquote{assertNotNull} but only the \enquote{Normal (Restricted)} pattern identifies the scenario as \enquote{protocol}.
Because of this difference in performance, it is important to order the patterns to produce the best results.

The ordering of both name and body patterns in our approach is based on our understanding of the patterns, the intuition that more specific patterns should be tried before more general patterns, and the results of applying them to the applications shown in \cref{tab:test-corpus} as a pilot study.
In this pilot study, we tested ten different arrangements of the patterns and selected the one that produced the most accurate ordering.
More details about this evaluation process can be find in \cref{sec:evaluation:feasibility,sec:evaluation:accuracy}.
The resulting orders for the name patterns and body patterns are shown in \cref{tab:name-patterns,tab:rq1-body}, respectively.

\subsection{Phase 2: Information Comparison}
\label{sec:InformationComparison}

The goal of the information comparison phase is to detect non-descriptive test names.
Our approach fulfills this goal by comparing the information extracted from the test name and body.
The result of this comparison is that a test name is either:
\begin{enumerate*}[itemjoin*={{, or }}]
    \item Descriptive
    \item Non-descriptive
    \item Unknown
\end{enumerate*}.

More specifically, each piece of information extracted from a test's name is compared with its corresponding piece of information extracted from the test's body (i.e., action\textsubscript{name} with action\textsubscript{body}, predicate\textsubscript{name} with predicate\textsubscript{body}, and scenario\textsubscript{name} with scenario\textsubscript{body}).
If the action, predicate, and scenario extracted from the name are all empty and\slash or the action, predicate, and scenario extracted from the body are all empty, the name is characterized as Unknown.
In this case, it is impossible to determine the quality of the name because an insufficient amount of information was extracted from the name or body.

\begin{figure}[t]
    \centering
    \begin{subfigure}{0.9\textwidth}
    \centering
        \includecode[java]{detected4.java}
        \caption{Example test with a descriptive name.}
        \label{PatternExample_detected4_code}
    \end{subfigure}
    \\[0.5ex]
    \begin{subfigure}{0.9\textwidth}
    \centering
        \begin{tabular}{rcc}
        \toprule
                  & Name    & Body    \\
        \midrule
        action    & GetGraphNode  & getGraphNode()     \\
        predicate & ---           & assertEquals()     \\
        scenario  & ---           & ---                \\
        \bottomrule
        \end{tabular}
        \caption{Extracted information for the test shown in \cref{PatternExample_detected4_code}.}
        \label{PatternExample_detected4}
    \end{subfigure}
    \caption{Example of a descriptive test name.}
    \label{fig:descriptive-examples}
\end{figure}

\begin{figure}[t]
    \centering
    \begin{subfigure}{0.8\textwidth}
    \centering
        \includecode[java]{detected1.java}
        \caption{Example test with a non-descriptive name.}
        \label{PatternExample_detectedPoorName_code}
    \end{subfigure}
    \\[0.5ex]
    \begin{subfigure}{0.9\textwidth}
    \centering
        \begin{tabular}{rcc}
        \toprule
                  & Name           & Body      \\
        \midrule
        action    & ---            & extract() \\
        predicate & ThrowException & ---       \\
        scenario  & TokenIsAbsent  & response  \\
        \bottomrule
        \end{tabular}
        \caption{Extracted information for the test shown in \cref{PatternExample_detectedPoorName_code}.}
        \label{PatternExample_detectedPoorName}
    \end{subfigure}
    \caption{Examples of a non-descriptive test name.}
    \label{fig:non-descriptive-examples}
\end{figure}

If there is sufficient information to compare, the approach checks each existing piece of information from the name against the corresponding information from the body.
If all of the existing pieces of information match, then the name is considered \emph{descriptive}.
Also, if all of the existing pieces from the name is a valid subset of the pieces from the body, the name is still considered \emph{descriptive}.
\Cref{fig:descriptive-examples} shows an example of a test name that is classified as descriptive.
The top of the figure shows the test under consideration and the bottom presents a table showing the information extracted by the first phase of the approach.
The rows of the table show the values extracted by the pattern type shown in the corresponding column (i.e., the name pattern identified \enquote{GetGraphNode} as the action and the body pattern identified \enquote{getGraphNode()} as the action).
In this example, the name is considered descriptive because all of the non-empty information types match their counterpart (i.e., \enquote{GetGraphNode} matches \enquote{getGraphNode()}).

If, when comparing the name information against the body information, at least one of the existing pieces of information does not match, then the name is considered \emph{non-descriptive}.
It means that a subset of the following conditions happens for that name:
\begin{enumerate*}[itemjoin*={{, or }}]
    \item action\textsubscript{name} does not match action\textsubscript{body}
    \item predicate\textsubscript{name} does not match predicate\textsubscript{body}
    \item scenario\textsubscript{name} does not match scenario\textsubscript{body}
\end{enumerate*}.
\Cref{fig:non-descriptive-examples} shows an example of name that is classified as non-descriptive.
Again, the top of the figure shows the test under consideration and the bottom presents a table showing the information extracted by the first phase of the approach.
In this example, the name is considered non-descriptive because none of the non-empty information types match their counterpart (i.e., \enquote{TokenIsAbsent} fails to match \enquote{response}).

If the outcome is either descriptive or non-descriptive, the approach can sometimes provide additional information to developers to help them improve the test name.
For both descriptive and non-descriptive names, if a value provided by the name pattern is empty but the corresponding information provided by the body is not empty, the name can likely be improved by the addition of the body information.
For example, \cref{PatternExample_detected4} shows a test name that is descriptive but can also be improved.
In this example, the name accurately reflects that the action of the test is \enquote{GetGraphNode} but it is missing information about the predicate that can be found in the body.
Adding information that the predicate is \enquote{assertEquals} to the name would improve its descriptiveness.

For only non-descriptive names, the approach can suggest modification in two cases.
First, if a value provided by the name pattern exists but the corresponding value provided by the body pattern does not exist, the approach suggests that the name information from the name be removed as it is not supported by the body.
Second, if the corresponding values provided by the name and body patterns both exist but do not match, the approach can suggest that the information from the name be replaced by the information from the body.
For example, \cref{PatternExample_detectedPoorName} shows a non-descriptive test name, which the approach can provide the following suggestions for improvement:
First, the predicate part of the name, \enquote{ThrowException}, should be removed and second, the scenario identified by the name, \enquote{TokenIsAbsent}, should also be replaced with the scenario from the body, \enquote{response}.
Note that, because the action from the name is empty, the action identified by the body, \enquote{extract}, should be added to the name, as described above.

The challenging part of this phase is determining whether the corresponding pieces of information match.
Because the information extracted from the name is text while the information extracted from the body are code elements (i.e., \texttt{methods}, \texttt{objects}, etc.) they can not be directly compared.
To address the challenge, the approach automatically converts the \emph{name} of any \texttt{method}, \texttt{object}, \texttt{new instance}, or \texttt{assertion method} to a string.
For example, the \enquote{Normal (Restricted)} body patterns can extract the name of the \texttt{assertion method} in \cref{fig:approach-0}, and it is converted to a string that is shown as \enquote{assertNotNull} in \cref{fig:approach-2}.
Once both the information from the name and the information from the body have been converted to strings, they are also converted to lower case.

The two strings are equal, or if one is strictly contained in the other (i.e., one of them may contain additional information), they are considered to match.
Otherwise, they are unmatched.

After we sorted out the process of determining a match, the pattern-based approach can automatically classify each test name in a project as a descriptive name or a non-descriptive name.
In the first step, the approach gathers all unit tests from the test suite of a selected project by using an automated project analyzer and finds pattern matches for their test names and bodies.
In the second step, the approach then uses the test patterns to extract the action, predicate, and scenario from the name and body of each test and generate a report for each name, which is the same as the reports shown in~\cref{NewExamplesForRQ3}.
In the last step, the approach automatically aggregates all generated reports for all extracted tests in a comprehensive report for the project, which contains all the detected descriptive and non-descriptive test names.
To provide a more intuitive presentation of the approach, an implementation of the pattern-based approach is provided as an IDE plugin~\cite{prototype}.

\section{Empirical Evaluation}
\label{sec:evaluation}

The overall goal of the evaluation is to determine if our approach can classify descriptive and non-descriptive test names. 
However, because the approach's success for this task depends on the underlying patterns, we also evaluate several aspects of their performance.
More specifically, we considered the following three research questions:
\begin{description}[font=\normalfont\emph]
\item[RQ1---Feasibility.] How many test names and bodies are matched by the patterns used by the approach?
\item[RQ2---Accuracy.] How accurate are the patterns at extracting the action, scenario, and predicate from test names and bodies?
\item[RQ3---Effectiveness.] Can our pattern-based approach correctly classify descriptive and non-descriptive test names?
\end{description}

To investigate these questions, we implemented our approach as an IntelliJ IDEA Plugin \cite{IntelliJPlugin}.
We chose to use IntelliJ IDEA because it is a full-featured IDE that can import projects which use a wide variety of build systems (e.g., Maven and Gradle).
This gives us more flexibility in choosing applications when building our set of experimental subjects.
It also has support for automatically identifying test suites, which are the input to the approach.
Finally, it has a robust parsing API that we can use to implement the body patterns.
Currently developers can use the plugin by selecting a menu item that analyzes all tests in the current project.
In future work, the plugin could easily be extended to support other interaction mechanisms.
For example, checking only the names of test in a specific class or the names of individually selected tests.

To generate the experimental data for investigating our research questions, we instrumented the plugin to record the information necessary for answering each research question.
We then manually ran the plugin on each of the experimental subjects.
For each run, the plugin automatically imports the project that is going to be evaluated.
After the importing finishes, the plugin will attempt to match every test pattern on each unit test that is contained in that imported project.
Finally, the plugin outputs a report for all unit tests that are evaluated in the process.
In total, we collected all information comparison reports for each of the ten Java projects we used in the evaluation.
The machine we used for all experiments was a MacBook Pro (\SI{2.7}{\giga\hertz} Intel i5 processor; 16 GB RAM) running macOS High Sierra and Java version 9.0.1.
Adding up the time for executing the plugin on each project, the total amount of time is roughly five hours for \num{34352} tests.
Even though the implementation is unoptimized, the execution time is such that it is feasible to include the approach as part of an off-line build process (e.g., overnight).

The remainder of this section describes our experimental subjects, research questions, and experimental results in more detail.

\subsection{Experimental Subjects}
\label{sec:evaluation:subjects}

\begin{table*}
\centering
\caption{Experimental Subjects.}
\begin{tabular}{
  l
  l
  S[table-format=5]
}
 \toprule 
 \multicolumn{1}{c}{\textbf{Project}} & \multicolumn{1}{c}{\textbf{Commit}} & \multicolumn{1}{c}{\textbf{\# Tests}} \\
 \midrule
 Xodus      & 8d82ef7 & 940   \\
 mytcuml    & 0786c55 & 21532 \\
 wheels     & 15696da & 811  \\
 EventBus   & 2e7c046 & 124  \\
 Picasso    & 5c05678 & 336  \\
 Jenkins    & 6c1d61a & 2245 \\
 ScribeJava & fce41f9 & 109  \\
 mockito    & 2204944 & 2112 \\
 Guice      & 6f1c6cc & 1322 \\
 fastjson   & 4c7935c & 4821 \\
 \midrule
 \multicolumn{2}{r}{Total} & 34352 \\
 \bottomrule
\end{tabular}
\label{tab:subjects}
\end{table*}

As the subjects for the evaluation, we selected a set of \num{34352} unit tests comprised of the test suites from the \num{10} Java projects shown in \cref{tab:subjects}.
In the table, the first column, \emph{Project}, shows the name of each project; the second column, \emph{Commit}, shows commit hash of the version of the project that was evaluated; and the last column, \emph{\#~Tests}, shows the number of unit tests contained in each project's test suite.

We chose these projects for several reasons.
First, they are distinct from the applications and test suites we used to develop the patterns (see \cref{sec:test_patterns}).
Clearly, the patterns should perform well on the tests that they were derived from.
Having separate test suites allows for a more representative evaluation of the first two research questions.
Second, the applications they test are diverse since they cover a wide variety of application domains.
For example, \enquote{Xodus} is a transactional database, \enquote{mytcuml} is a UML tool, \enquote{wheels} is a testing framework, and \enquote{EventBus} is a publish\slash subscribe pattern-based library that can simplify Android and Java code.
In addition, they were written by different developers and at different times.
This means that the test suites are not limited to a particular set of authors or patterns and are more likely to be representative than any test from a single project or style.
Finally, in aggregate, they have a sufficient number of tests to allow for a thorough evaluation of the approach.

\subsection{RQ1: Feasibility}
\label{sec:evaluation:feasibility}

\begin{table*}[t]
\centering
\caption{Match Rates for Name Patterns (Over All Tests).}
\begin{tabular}{
  l
  S[table-format=5]
  S[table-format=3.2]
}
 \toprule 
 \multicolumn{1}{c}{\textbf{Name Pattern}} & \multicolumn{1}{c}{\textbf{\# Matches}} & \multicolumn{1}{c}{\textbf{(\%)}} \\
 \midrule
 Verb With Multiple Nouns Phrase           & 0                                       & 0.00  \\
 Divided Duel Verb Phrase                  & 0                                       & 0.00  \\
 Is And Past Participle Phrase             & 0                                       & 0.00  \\
 Try Catch                                 & 204                                     & 0.59  \\
 Duel Verb Phrase                          & 331                                     & 0.96  \\
 Noun Phrase                               & 1555                                    & 4.53  \\
 Single Entity                             & 4794                                    & 13.96 \\
 Verb Phrase Without Prepended Test        & 2578                                    & 7.50  \\
 Verb Phrase With Prepended Test           & 9007                                    & 26.22 \\
 Regex Match                               & 15883                                   & 46.24 \\
 \midrule
 Overall                                   & 34352                                   & 100.00 \\
 \bottomrule
\end{tabular}
\label{tab:rq1-name}
\end{table*}

\begin{table*}[t]
\centering
\caption{Match Rates for Body Patterns (Over All Tests).}
\begin{tabular}{
  l
  S[table-format=5]
  S[table-format=3.2]
}
 \toprule 
 \multicolumn{1}{c}{\textbf{Body Pattern}} & \multicolumn{1}{c}{\textbf{\# Matches}} & \multicolumn{1}{c}{\textbf{(\%)}} \\
 \midrule
 If Else                                   & 17                                      & 0.05   \\
 Loop                                      & 533                                     & 1.55   \\
 All Assertion                             & 1801                                    & 5.24   \\
 No Assertion                              & 3602                                    & 10.49  \\
 Try Catch                                 & 5075                                    & 14.77  \\
 Normal (Restricted)                       & 1634                                    & 4.76   \\
 Normal (Generalized)                      & 13840                                   & 40.29  \\
 \midrule
 Overall                                   & 26502                                   & 77.15  \\
 \bottomrule
\end{tabular}
\label{tab:rq1-body}
\end{table*}

The purpose of the first research question is to evaluate the feasibility of our pattern-based approach.
The primary way in which we judge feasibility is to determine the percentage of test names and bodies that are matched by one of the patterns used by the approach.
In some sense, this is the \enquote{coverage} of the patterns.
If the coverage of the patterns is low, the usefulness of the approach will also be low as the approach will only be able to provide feedback for a small number of tests.
Conversely, if the coverage of the patterns is high, the approach is potentially more useful as it can provide feedback for more tests.
However, there is a potential trade-off between the coverage of the patterns and their accuracy (see \cref{sec:evaluation:accuracy}) in that increasing coverage may result in lower accuracy.
As such, the sweet-spot for the approach is achieving enough coverage to enable providing feedback for most tests, but not compromising the accuracy of the extracted information.

\Cref{tab:rq1-name,tab:rq1-body} show the experimental data for this research question.
In each table, the first column, \emph{Name\slash Body Pattern}, shows the name of each pattern; the final two columns, \emph{\# Matches} and \emph{\%}, show the number of times the pattern matched against a test both as a count and a percentage, respectively; and the final row shows an overall summary of the results.
For example, the fourth row of \cref{tab:rq1-name} shows that \enquote{Try Catch} matched \num{204} test names (i.e., \SI{\approx 0.6}{\percent} of the \num{34352} considered test names).
Similarly, the first row of \cref{tab:rq1-body} shows that \enquote{If Else} matched \num{17} of the \num{34352} considered test bodies (i.e., \SI{\approx 0.05}{\percent} of the \num{34352} considered test bodies).
Note that, to simplify the tables and the discussion, most variations of a pattern are grouped into a single row.  For example, in \cref{tab:rq1-body}, \enquote{All Assertion} includes both the \enquote{Single} and the \enquote{Multiple} versions presented in \cref{tab:body-patterns}.

As the final row in each table shows, the overall match rate for both name and body patterns is high.
In aggregate, the name patterns matched \num{34352} test names (i.e., \SI{100}{\percent}), and the body patterns matched \num{26502} test bodies (i.e., \SI{\approx 77}{\percent}).
While there are a few patterns that had low or zero match rates (e.g., \enquote{Divided Duel Verb Phrase}), the cost of keeping such patterns is low as their execution times are low and they may be more prevalent in other project types.
The data also demonstrate that the ordering of the patterns is effective.
More general patterns (i.e., ones have shown lower in the tables) have higher match rates than more specific patterns (ones shown higher in the tables).
Overall, we believe that these results suggest that the approach is feasible.
Based on the performance of several related approaches~\cite{host2009debugging,host2008java,zhong2013detecting,singer2008exploiting,allamanis2014learning}, we believe that the coverage of the patterns is high enough to enable the approach to provide feedback for a majority of tests.

\subsection{RQ2: Accuracy}
\label{sec:evaluation:accuracy}

\begin{table*}[t]
\caption{Accuracy Results for Each Name Pattern.}
\begin{adjustbox}{max width=\textwidth}
\begin{tabular}{
l
S[table-format=3.1]
S[table-format=3.1]
S[table-format=3.1]
S[table-format=3.1]
S[table-format=3.1]
S[table-format=3.1]
}
 \toprule 

 &  \multicolumn{2}{c}{\textbf{Action (\%)}}  &  \multicolumn{2}{c}{\textbf{Predicate (\%)}}  & \multicolumn{2}{c}{\textbf{Scenario (\%)}} \\
 
 \cmidrule(lr){2-3} \cmidrule(lr){4-5} \cmidrule(lr){6-7}
 
\multicolumn{1}{c}{\textbf{Name Pattern}} & \multicolumn{1}{c}{\textbf{TP}} & \multicolumn{1}{c}{\textbf{FP}} & \multicolumn{1}{c}{\textbf{TP}} & \multicolumn{1}{c}{\textbf{FP}} & \multicolumn{1}{c}{\textbf{TP}} & \multicolumn{1}{c}{\textbf{FP}} \\
 \midrule
  Verb With Multiple Nouns Phrase    & {---} & {---} & {---} & {---} & {---} & {---} \\
  Divided Duel Verb Phrase           & {---} & {---} & {---} & {---} & {---} & {---} \\
  Is And Past Participle Phrase      & {---} & {---} & {---} & {---} & {---} & {---} \\
  Try Catch                          & 89  & 11 & 96  & 4  & 89  & 11                \\
  Duel Verb Phrase                   & 96  & 4  & 88  & 12 & 84  & 16                \\
  Noun Phrase                        & 100 & 0  & 100 & 0  & 100 & 0                 \\
  Single Entity                      & 97  & 3  & 97  & 3  & 89  & 11                \\
  Verb Phrase With Prepended Test    & 87  & 13 & 74  & 26 & 95  & 5                 \\
  Verb Phrase Without Prepended Test & 100 & 0  & 75  & 25 & 75  & 25                \\
  Regex Match                        & 84  & 16 & 84  & 16 & 72  & 28                \\
  \midrule
  Overall                            & 92  & 8  & 87  & 13 & 89  & 11                \\
 \bottomrule
\end{tabular}
\end{adjustbox}
\label{tab:rq2_name}
\end{table*}

\begin{table*}[t]
\centering
\caption{Accuracy Results for Each Body Pattern.}
\begin{tabular}{
  l
  S[table-format=3.1]
  S[table-format=3.1]
  S[table-format=3.1]
  S[table-format=3.1]
  S[table-format=3.1]
  S[table-format=3.1]
}
\toprule 
 &  \multicolumn{2}{c}{\textbf{Action (\%)}}  &  \multicolumn{2}{c}{\textbf{Predicate (\%)}}  & \multicolumn{2}{c}{\textbf{Scenario (\%)}} \\
 
 \cmidrule(lr){2-3} \cmidrule(lr){4-5} \cmidrule(lr){6-7}
 
\multicolumn{1}{c}{\textbf{Body Pattern}} & \multicolumn{1}{c}{\textbf{TP}} & \multicolumn{1}{c}{\textbf{FP}} & \multicolumn{1}{c}{\textbf{TP}} & \multicolumn{1}{c}{\textbf{FP}} & \multicolumn{1}{c}{\textbf{TP}} & \multicolumn{1}{c}{\textbf{FP}} \\
 \midrule
  If Else              & 91  & 9  & 36  & 64 & 100 & 0 \\
  Loop                 & 89  & 11 & 86  & 14 & 94  & 6 \\
  All Assertion        & 100 & 0  & 89  & 11  & 100 & 0 \\
  No Assertion         & 96  & 4  & 74  & 26 & 100 & 0 \\
  Try Catch            & 100 & 0  & 94  & 6  & 91  & 9 \\
  Normal (Restricted)  & 100 & 0  & 100 & 0  & 100 & 0 \\
  Normal (Generalized) & 82  & 18 & 100 & 0  & 96  & 4 \\
  \midrule
  Overall              & 94  & 6  & 88  & 12 & 97  & 3 \\
 \bottomrule
\end{tabular}
\label{tab:rq2_body}
\end{table*}

The goal of the second research question is to investigate whether the patterns can accurately extract information from test names and bodies.
Because assessing the accuracy of the extracted information must be done manually (i.e., inspect each test case with the information manually and check if it can correctly describe the action\slash predicate\slash scenario of the name or body by our researchers), it is infeasible to consider all \num{26502} tests that were matched by a pattern.
Therefore, we chose a subset of information extracted from matched tests to classify.
For each name pattern and each body pattern, we randomly selected up to \num{5} tests matched by that pattern from each project.
If no test was matched by that pattern in a project, we skipped the project and moved on to the next one.
In total, \num{242} tests were selected for the name patterns and \num{266} tests were selected for the body patterns.

For each test in the selected subset, each author manually examined the information extracted by the matching name and body patterns independently.
If the extracted information matched the human's judgment it was considered a true positive (TP) and if the extracted information did not match the human's judgment it was considered a false positive (FP).
Disagreements among the raters were discussed until a resolution was reached.
In total, \num{1524} comparisons were made by each rater (i.e., (\num{242} tests for name patterns + \num{266} tests for body patterns) * \num{3} comparisons, the action, predicate, and scenario for each test).

\Cref{tab:rq2_name,tab:rq2_body} show the experimental data for this research question.
\Cref{tab:rq2_name} shows the accuracy of the name patterns and \cref{tab:rq2_body} shows the accuracy of the body patterns.
In each table, the first column is the name of each pattern, and the following three pairs of columns show the TP and FP rates for the information extracted as the action, predicate, and scenario by each pattern.
The final row shows the overall rates for all patterns.
For example, the fourth row of \cref{tab:rq2_name} shows the accuracy results for the \enquote{Try Catch} name pattern: the TP rate for the action is \SI{89}{\percent}, the TP rate for the predicate is \SI{96}{\percent}, and the TP rate for the scenario is \SI{89}{\percent}.
Note that in \cref{tab:rq2_name} a dash (---) indicates the cases where a manual assessment was impossible because the patterns did not match any tests.

The data shown in \cref{tab:rq2_name} and \cref{tab:rq2_body}, indicates that the overall accuracy of both the name patterns and body patterns is high.
For name patterns, the overall true positive rates range from \SI{87}{\percent} for the scenario to \SI{92}{\percent} for the action and for the body patterns the overall true positive rates range from \SI{88}{\percent} for the predicate to \SI{97}{\percent} for the scenario.
Even in the worst cases (e.g., identifying the scenario with the Regex Match name pattern), the true positive rate is above \SI{70}{\percent}.
As such, we believe that both types of patterns are effective at accurately identifying the action, predicate, and scenario from tests.

\subsection{RQ3: Effectiveness}

\begin{table*}[t]
\centering
\caption{Effectiveness of the approach.}
\begin{tabular}
{
  l
  S[table-format=3.1]
  S[table-format=3.1]
  S[table-format=3.1]
  S[table-format=3.1]
}
\toprule
& & \multicolumn{2}{c}{\textbf{Rate (\%)}}\\
\cmidrule(lr){3-4}

\multicolumn{1}{c}{\textbf{Project}} &
\multicolumn{1}{c}{\textbf{\# Reports}} & 
\multicolumn{1}{c}{\textbf{TP}} &
\multicolumn{1}{c}{\textbf{FP}}
\\
\midrule
 Xodus      & 29  &  97  & 3   \\
 mytcuml    & 105 &  96  & 4   \\
 wheels     & 11  &  91  & 9   \\
 EventBus   & 10  &  90  & 10  \\
 Picasso    & 14  &  93  & 7   \\
 Jenkins    & 20  &  90  & 10  \\
 ScribeJava & 3   &  100 & 0   \\
 mockito    & 11  &  82  & 18  \\
 Guice      & 16  &  94  & 6   \\
 fastjson   & 46  &  98  & 2   \\
 \midrule
 Overall    & 265 &  95  &  5  \\
\bottomrule
\end{tabular}
\label{tab:rq3}
\end{table*}

The goal of the third research question is to determine if the pattern-based approach can correctly classify descriptive and non-descriptive test names.
Like for RQ2, assessing the output of the approach is a manual process that can not be applied to every output.
Therefore, we again selected a representative subset to consider.
In this case, because we are interested in the performance across all tests, we chose to consider a total of  \num{265} tests (i.e., \SI{1}{\percent} of the \num{26502} tests matched by both a name and body pattern).
The \num{265} tests were selected from among each project proportionally to the number of tests in the project's test suite (e.g., \num{105} tests were taken from \enquote{mytcuml}, \num{46} test were taken from \enquote{fastjson}, etc.).

For each test in the selected subset, each author again manually examined the output of the approach independently.
If the output of the approach matched the human's judgment it was considered a true positive (TP) and if the output did not match the human's judgment it was considered a false positive (FP).
Disagreements among the raters were discussed until a resolution was reached.
The results of the classification are shown in \cref{tab:rq3}.

In \cref{tab:rq3}, the first column presents each project’s name, the second column shows the number of tests for each project, and the final two columns show the rates for each classification, respectively.
For example, the first row shows there are \num{29} reports that were selected for project \enquote{Xodus}, and (\SI{\approx 97}{\percent}) of them correctly classify the test name as either descriptive or non-descriptive.
The last row shows that the overall TP rate of all reports is \SI{\approx 95}{\percent} (i.e., \num{251} true positives), and the FP rate is just \SI{\approx 5}{\percent} (i.e., \num{14} false positives).
Owing to the high effectiveness of our pattern-based approach, \SI{\approx 99.2}{\percent} of the \num{251} true positives are definitively correct.
And the definition of correctness here is to be a suitable test name for its related test body.
For instance, the test case in \cref{fig:descriptive-examples} is considered to be a true positive since the report can correctly classify its name as a descriptive test name.
Additionally, more examples of true positives can be found in the public repository~\cite{prototype}.
Because the true positive rate is high with nearly perfect correctness rate, we can conclude that the pattern-based approach is effective at classifying names as either descriptive or non-descriptive.

\begin{figure}[t]
    \centering
    \begin{subfigure}{0.75\textwidth}
    \centering
        \begin{footnotesize}
\begin{verbatim}
CovariantOverrideTest.returnFoo2
Action:    n/a != return
Predicate: assertEquals != n/a
Scenario:  thenReturn != foo2
-> Non-descriptive
\end{verbatim}
        \end{footnotesize}
        \caption{A correctly detected non-descriptive test name.}
        \label{NewExamples3-1}
    \end{subfigure}
    \\[2ex]
    \begin{subfigure}{0.75\textwidth}
    \centering
        \begin{footnotesize}
\begin{verbatim}
SmartNullsStubbingTest.shouldNotThrowSmartNullPointerOnObjectMethods
Action:    tostring != null
Predicate: n/a != null
Scenario:  null == null
-> Non-descriptive
\end{verbatim}
        \end{footnotesize}        
        \caption{A wrongly detected non-descriptive test name.}
        \label{NewExamples3-2}
    \end{subfigure}
    \caption{Examples of a true positive and a false positive.}
    \label{NewExamplesForRQ3}
\end{figure}

In addition, we further investigate the two examples in~\cref{NewExamplesForRQ3} that are selected from the \num{265} tests from those open-source Java projects.
In~\cref{NewExamplesForRQ3}, each example is the output that is produced by our pattern-based approach.
The action, predicate, and scenario on the left side of the equations are extracted from the test body, and the action, predicate, and scenario on the right side are extracted from the test name.
For the unit test in \cref{NewExamples3-1}, the test name \texttt{returnFoo2} is correctly classified as a non-descriptive test name.
Also, some suggestions are provided by our approach for the example in~\cref{NewExamples3-1}: 
\begin{enumerate*}
    \item the action of the name (i.e., \texttt{return}) should be removed from the name
    \item the predicate and scenario of the name should be replaced by the predicate and scenario of the body (i.e., \texttt{equals} and \texttt{thenReturn})
\end{enumerate*}.
For the unit test in \cref{NewExamples3-2}, the test name \texttt{shouldNotThrowSmartNullPointerOnObjectMethods} is incorrectly classified as a non-descriptive test name.
Because of the difference in length between the short name and long body, the test patterns failed to correctly extract the action, predicate, and scenario from the name and body, and this example is also considered as a false positive.

\section{Related Work}
 
In this paper, we propose a pattern-based approach that involves different fields of research, so the purpose of this section is to review the most closely related works that come from each field.

\subsection{Detecting Mismatches\slash Improving Names}

There are some existing works that attempt to identify name\slash implementation mismatches.

\Citeauthor{host2009debugging}'s work is the most similar to our approach as it attempts to identify several types of naming bugs in general Java methods~\cite{host2009debugging}.
Their approach relies on a manually generated rule book that is extracted from the implicit convention between names and implementations in Java programming, which can be utilized to detect name bugs and provide some suggestions for constructing more suitable names.
In their previous works, \citeauthor{host2008java} already showed that there is a mutual dependency between method names and their associated implementations~\cite{host2008java}.
Therefore, their approach considered the mismatch between the name and the implementation of its associated method and used the mismatch to fix name bugs, which are both similar to the analytical process and goal of our pattern-based approach.
There are two major differences between their work and ours.
First, our approach primarily focuses on the test names rather than general method names that often follow a different naming convention.
For example, their approach treated the data type of the value in the \texttt{return} statement as an essential attribute in their rule book.
However, normally for the unit tests, they compared different values using the \texttt{assertions} rather than any \texttt{return} statement, so the information in those \texttt{assertions} will be a crucial part of their test names.
Second, instead of using a manually generated rule book, we built our approach based on the test patterns, and those test patterns were mined from a large test corpus by a semi-automatic process.

\citeauthor{zhong2013detecting} provided a novel approach for detecting API documentation errors, and those errors are essentially the mismatches between the API documentation and the actual programs~\cite{zhong2013detecting}.
To address the importance of words in Java programming, \citeauthor{singer2008exploiting} showed that words in class names are closely related to class properties that can be described in micro patterns~\cite{singer2008exploiting}.
\citeauthor{allamanis2014learning} mentioned that developers should follow a consistent naming convention, and they proposed a novel framework that can suggest identifier names accurately~\cite{allamanis2014learning}.
All of their works comprehensively showed it is feasible to find poorly structured (i.e., non-descriptive) names by using the mismatch or pattern between the name and the program, and we can also improve those names by using providing accurate suggestions.
Nonetheless, each of their techniques is often limited to a certain aspect in the problem of detecting and improving non-descriptive names, so none of them can be directly applied to improving non-descriptive names in unit tests.
\citeauthor{pradel2018deepbugs} recently proposed a framework for the detection of naming bugs~\cite{pradel2018deepbugs}.
Regardless of their effort to introduce a new approach that can detect name-based bugs by using their machine learning method, we still can not apply their approach to the unit tests without further modifications.
Because some unit tests are expected to produce certain exceptions or failures when using them, so testers might intentionally design poorly named identifiers in those tests.
Consequently, lots of false-positives could be generated without a complete retrofit to extend their proposed framework to unit tests.
For instance, Junit 4 requires every test name to have a leading \enquote{test} \cite{JUnit4}, so some of existing techniques might consider the leading \enquote{test} as the action, predicate, or scenario of the name.

\subsection{Automated Generation of Test Names}

In contrast to the techniques mentioned above that attempt to improve names, there are several approaches that attempt to automatically generate names.

Some of these techniques use natural language-based program analysis~(NLPA).
For example, \citeauthor{zhang2016towards} proposed their approach that can generate descriptive names from existing test bodies by using natural-language program analysis and text generation~\cite{zhang2016towards}.
However, their approach left an important question unanswered that is testers need to decide which one of the three possible test names should be used for their unit tests by themselves, and it is possible that none of the three generated names follow the common naming convention.
Other techniques utilized Java bytecode, method-call sequences, API-level coverage goals, and \texttt{logbilinear} context models~\cite{fraser2011evosuite,thummalapenta2009mseqgen,daka2017generating,allamanis2015suggesting}.
Even with their automated generation process, their generated test names are not human-readable that can cause misunderstanding for testers who want to further modify those generated test names or bodies.
Although some techniques can generate descriptive names, those techniques required testers to perform a full test execution with certain coverage goals or building a context model, which are often error-prone in practice (i.e., those coverage goals or models might be too specific to apply for certain projects).

\subsection{General Program Analysis\slash Automated Testing and Debugging}

Many researchers proposed their program analysis or automated testing techniques that can help us have a better understanding of the embedded features in unit tests. 

\citeauthor{moreno2012jstereocode} proposed Java method and class stereotypes, and they took a closer look at the statement level analysis of Java code~\cite{moreno2012jstereocode}, and~\citeauthor{ghafari2015automatically} tried to extract the focal method under test~\cite{ghafari2015automatically}.
To be more focused on unit testing,~\citeauthor{li2018aiding} constructed a series of tags for distinguish unit test cases~\cite{li2018aiding}.
A group of researchers also conducted a series of works related to tagging methods or classes with stereotypes~\cite{dragan2006reverse,dragan2010automatic,dragan2011emergent}, but their works might also not be applicable for unit tests.
From a general perspective of testing, other researchers tried to devise methods that can perform fully automated testing by the targeted event sequence or the environmental dependencies~\cite{jensen2013automated,arcuri2014automated}.
All of their works performed well under their specific problems in program analysis or automated testing. 
However, while their works focus more on extracting features from code or automating the testing process rather than the unit tests themselves, we can still use them to improve our pattern-based approach.
Furthermore, \citeauthor{li2019deepfl} proposed a learning-based approach for fault localization and automated debugging with high performance~\cite{li2019deepfl}, but the goal of their work is primarily to locate software faults for debugging rather than the naming of unit tests. 
Regardless of the performance of their proposed tool, the goal of their work is primarily to locate software faults for debugging rather than the naming of unit testing, and the experimental subjects they used is a standard benchmark database of detecting bugs rather than the unit tests from real-world Java projects.

\subsection{Natural Language Program Analysis}

There are lots of existing works that try to analyze programs from a natural language-based perspective.

\citeauthor{pollock2007introducing} and~\citeauthor{shepherd2007using} introduced NLPA by illustrating how to apply NLPA in practice and also giving some insights about aspect mining~\cite{pollock2007introducing,shepherd2007using,pollock2009natural}.
Their studies showed natural language clues from developers' naming style can be used for improving the effectiveness of software tools.
\citeauthor{abebe2010natural} proposed a natural language-based method to parse the identifier names of program elements for extracting concepts from them~\cite{abebe2010natural}.
Furthermore, some researchers attempted to split identifiers~\cite{enslen2009mining,butler2011improving,guerrouj2013tidier,hill2014empirical}, and others managed to expand abbreviations~\cite{hill2008amap,madani2010recognizing,corazza2012linsen}.
Even though their works are not alternatives to our approach, we can still use their implemented tools to improve the accuracy of the test patterns.

\section{Prototype Implementation and Threats to Validity}
\label{sec:implementation&threats}

The prototype implementation is publicly available~\cite{prototype}.
All meta results from the pilot study and the pattern mining process, all the instances of non-descriptive test names from the 10 experimental subjects in~\cref{tab:subjects}, and the metadata of the evaluation are also uploaded in the repository.
In addition, we are sharing data for the quantitative analysis that was performed in the evaluation~\cite{evaluation_data}.
Two threats to validity do exist for our test name\slash body patterns:
\begin{enumerate*}
\item some body patterns contain a type of statement that can be cryptically constructed
\item some name patterns do not currently have a match
\end{enumerate*}.
For the first part, although it is rare to have cryptically constructed statements in a test body, we mitigated it by supporting those cryptically constructed statements within their corresponding patterns.
For example, we provided support for the conditional expression in the \texttt{If Else} body pattern to handle the cryptically constructed if else statement (e.g.,~\texttt{expression1?expression2:expression3;}).
For the second part, we will further conduct a large-scale empirical evaluation on at least \num{100} Java projects in order to discover potential matches for those name patterns as part of our planned future work.

\section{Conclusions and Future Work}

Taking every test pattern into consideration, our selected test patterns can extract sufficient information from any unit test with matched name\slash body patterns.
With the help of the output generated by our approach, developers can easily find non-descriptive test names from a given test corpus and improve those non-descriptive names by referring to the descriptive information.
Furthermore, we also implemented our approach as an IntelliJ IDE plugin.
In the empirical evaluation, the experimental results produced by our implemented approach are encouraging, which show our approach not only can accurately extract descriptive information from unit tests but also can correctly classify descriptive and non-descriptive test names.

For our planned future work, one possible direction is constructing an advanced version of the information comparison to improve the pattern-based approach by using more sophisticated comparing criteria.
The next step should be looking into the false-positives in the evaluation to see if we can further improve existing test patterns.
For example, we can further improve some name patterns by using even more accurate POS tagging or extend certain body patterns to handle different coding styles.
The last step of this direction is to conduct another large-scale evaluation with at least \num{100} Java projects from Github as experimental subjects.
To expand the scope of test patterns, an empirical study will be performed on other unit testing frameworks like \texttt{csUnit} and \texttt{PyUnit}, which are designed for \texttt{C\#} and \texttt{Python}.
Using the outcome of the large-scale study, we can determine if it is possible to mine and extract similar patterns from other types of unit tests and whether it might also be feasible to use mined patterns to extend the pattern-based approach to testing framework and programming languages.

\section{Acknowledgments}

This work is supported in part by National Science Foundation Grant\-No.~1527093.

\bibliography{paper.bib}

\end{document}